\documentclass[
 amsmath,amssymb,
 reprint,
author-numerical,%
]{revtex4-1}

\usepackage{graphicx}
\usepackage{dcolumn}
\usepackage{bm}

\usepackage[utf8]{inputenc}
\usepackage[T1]{fontenc}
\usepackage{mathptmx}
\usepackage{etoolbox}
\usepackage{todonotes}
\usepackage{soul}
\usepackage{hyperref}

\makeatletter
\def\@email#1#2{%
 \endgroup
 \patchcmd{\titleblock@produce}
  {\frontmatter@RRAPformat}
  {\frontmatter@RRAPformat{\produce@RRAP{*#1\href{mailto:#2}{#2}}}\frontmatter@RRAPformat}
  {}{}
}%
\makeatother
\begin{document}


\title[]{Time-walk and jitter correction in SNSPDs at high count rates}
\author{Andrew Mueller$^*$}
 \email{amueller@caltech.edu}
 \affiliation{Applied Physics, California Institute of Technology, 1200 E California Blvd, Pasadena, CA, USA}
 \affiliation{Jet Propulsion Laboratory, California Institute of Technology, 4800 Oak Grove Dr., Pasadena, CA, USA}

\author{Emma E. Wollman}
\affiliation{Jet Propulsion Laboratory, California Institute of Technology, 4800 Oak Grove Dr., Pasadena, CA, USA}

\author{Boris Korzh}
\affiliation{Jet Propulsion Laboratory, California Institute of Technology, 4800 Oak Grove Dr., Pasadena, CA, USA}

\author{Andrew D. Beyer}
\affiliation{Jet Propulsion Laboratory, California Institute of Technology, 4800 Oak Grove Dr., Pasadena, CA, USA}

\author{Lautaro Narvaez}
\affiliation{Division of Physics, Mathematics and Astronomy, California Institute of Technology, 1200 E California Blvd., Pasadena, CA 91125, USA}

\author{Ryan Rogalin}
\affiliation{Jet Propulsion Laboratory, California Institute of Technology, 4800 Oak Grove Dr., Pasadena, CA, USA}

\author{Maria Spiropulu}
\affiliation{Division of Physics, Mathematics and Astronomy, California Institute of Technology, 1200 E California Blvd., Pasadena, CA 91125, USA}
 
\author{Matthew D. Shaw}
\affiliation{Jet Propulsion Laboratory, California Institute of Technology, 4800 Oak Grove Dr., Pasadena, CA, USA}

\date{\today}

\begin{abstract}
Superconducting nanowire single-photon detectors (SNSPDs) are a leading detector type for time correlated single photon counting, especially in the near-infrared. When operated at high count rates, SNSPDs exhibit increased timing jitter caused by internal device properties and features of the RF amplification chain. Variations in RF pulse height and shape lead to variations in the latency of timing measurements. To compensate for this, we demonstrate a calibration method that correlates delays in detection events with the time elapsed between pulses. The increase in jitter at high rates can be largely canceled in software by applying corrections derived from the calibration process. We demonstrate our method with a single-pixel tungsten silicide SNSPD and show it decreases high count rate jitter. The technique is especially effective at removing a long tail that appears in the instrument response function at high count rates. At a count rate of 11.4 MCounts/s we reduce the full width at one percent maximum level (FW1\%M) by 45\%. The method therefore enables certain quantum communication protocols that are rate-limited by the (FW1\%M) metric to operate almost twice as fast. \\ \\
© 2022. All rights reserved.
\end{abstract}

\maketitle

Over the last decade, SNSPDs have advanced rapidly to become essential components in many optical systems and technologies, owing to their high efficiency $(>90\%)$~\cite{99.5_Chang_2021, reddy2020superconducting}, fast reset times ($< 1~\mathrm{ns}$)~\cite{Vetter2016Cavity} and scalability to kilopixel arrays~\cite{Wollman2019}. The timing jitter of SNSPDs is also best-in-class -- values as low as 3~ps have been demonstrated in short nanowires~\cite{Korzh2020}, and new high-efficiency designs exhibit sub-10 ps jitter~\cite{EsmaeilZadeh2020, Colangelo2021}.

\begin{figure}[!htbp]
  \centering
  \includegraphics[width=1\linewidth]{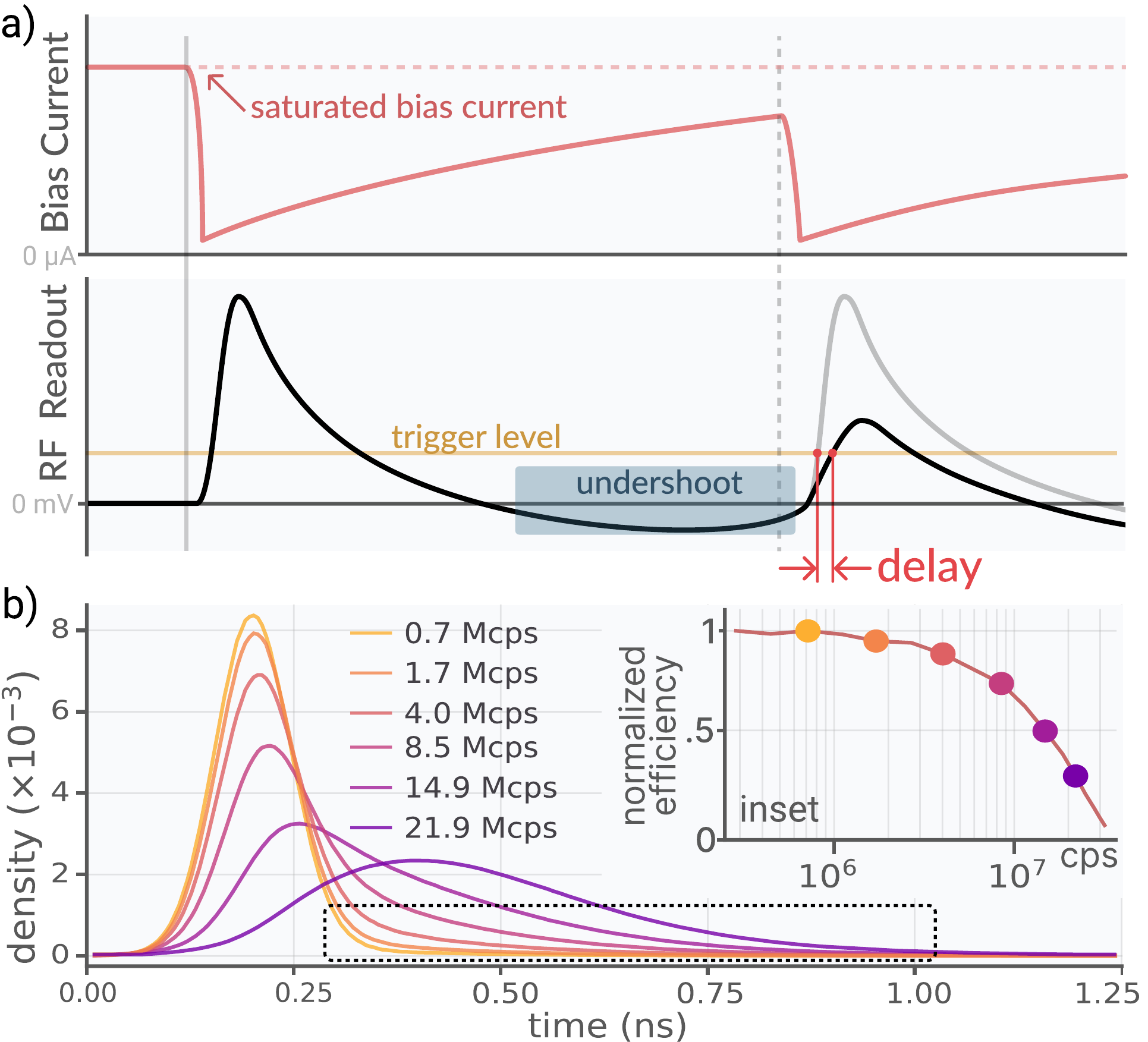}
  \caption{
    a) Diagram illustrating two major sources of correlated high count rate jitter. First, detections may occur during the reset time of a previous detection. At this time the bias current in the nanowire is below its saturated value so that a photodetection triggers an RF pulse with correspondingly lower amplitude. Second, an RF pulse may arrive in the undershoot region of a previous pulse, where the undershoot is a period of negative voltage induced by the low-frequency cutoff properties of the readout amplifier chain. b) Measured histograms of detections from short $3~\mathrm{ps}$ mode-locked laser pulses. With lower attenuation and higher count rate, the observed timing uncertainty is greater. Inset shows where respective count rates fall on a maximum count rate (MCR) curve \cite{Zhang_MCR_2019}. The MCR of an SNSPD is often quoted at the $3~\mathrm{dB}$ point, where the normalized efficiency drops to $-3~\mathrm{dB}$ of its maximum value. The jitter increase due to time walk manifests as a tail in the instrument response functions (inside dashed black box) even at count rates significantly below the 3-dB point where detector efficiency has not started to drop significantly (e.g. the 1.7 Mcps data). 
  }

  \label{fig:intro}
\end{figure}

SNSPD jitter increases with count rate due to properties of the nanowire reset process and features of the readout circuit. 
The effect bears resemblance to time walk observed in silicon avalanche diodes and other detectors where the pulses have varying heights and slew rates \cite{SPAD_walk_Kirchnir_1997} thereby causing a timing measurement using a fixed threshold to 'walk' along the rising edge of the pulse (the labeled delay in Fig. \ref{fig:intro}a). 
At low count rates SNSPDs exhibit very uniform pulse heights. However, at high counts rates where the inter-arrival time is on the order of the reset time of the detector, current-reset and amplifier effects lead to smaller and distorted pulses. If photon inter-arrival times are not known a priori in the intended application, the uncorrected time walk manifests as a perceived increase in jitter (Fig. \ref{fig:intro}b). 

The time-walk effect in SNSPDs is typically not reported, as jitter is usually measured at low count rates where the detector has ample time to reset following each detection.
But as communication and quantum information applications push into higher count rates, the high count rate induced jitter becomes more relevant. LIDAR, quantum and classical optical communication, and imaging applications may all benefit from the development of new detection systems and methods that keep jitter as low as possible in this regime. 

We first consider the features of SNSPD operation and readout that cause an increase in jitter with count rate. Then we present multiple ways of mitigating or avoiding these effects, before reviewing our preferred method that relies on a calibration and correction process. 

The jitter increase observed at high rate originates from two groups of system characteristics: (i) the intrinsic reset properties of the nanowire, and (ii) properties of the amplification chain. These influence the system's jitter differently, thus it is helpful to consider them separately. Added jitter from either or both sources can emerge when an SNSPD system is operated at high count rates.

The nanowire reset process determines how the detector becomes single-photon sensitive again after a detection. When an SNSPD fires, the current flow in the nanowire momentarily drops to near zero, then recovers in an inverted exponential fashion (Fig. \ref{fig:intro}a). An incident photon may trigger another detection before the bias current fully returns to its saturated value, producing a pulse with a lower amplitude and slew rate. A fixed threshold comparator will trigger on this lower pulse later in time than a full amplitude pulse. If uncorrected for, this time walk manifests as an increase in jitter at high count rates.

The choice of readout amplifiers may also contribute to added high count rate jitter. Pulses may be shifted in voltage and distorted if they arrive when the RF signal has not yet reached a steady zero state following the previous detection. For example, pulses may arrive within an amplifier-induced undershoot region following previous pulses. This phenomenon is illustrated in Fig. \ref{fig:intro}a as the area below 0~mV. The duration of ringing or undershoot effects varies widely depending on the design of the readout circuit. If they last longer than the reset time of the nanowire bias current, the calibration technique may correct for the potentially complex interactions between pulse waveforms that overlap in time.

There are various methods for correcting for increased jitter at high count rates. These include (i) the use of extra hardware that cancels out some distortions, or (ii) simple software-based data filtering that ignores distorted time tags. We review these techniques before covering the calibration and correction approach.

Variations in pulse height are a primary component of the distortions that appear at high rates. Such variations in other systems are commonly corrected with a constant fraction discriminator (CFD) that allows for triggering at a fixed percentage of pulse height rather than at a fixed voltage. Adding a CFD to an existing setup is straightforward, and there is precedent for their use with SNSPDs~\cite{Korzh2020}. But CFDs do not optimally correct for variations in pulse shape that go beyond varied vertical scaling. Also, multi-channel time-to-digital converters (TDCs) used to read out large SNSPD arrays typically only include fixed-threshold comparators~\cite{Wollman2019}. Implementing CFDs on many channels may not be straightforward or cost-effective.

In a simple software-based jitter mitigation method, each time-tagged event may be accepted or rejected based on how soon it arrives after the previous pulse. Those that arrive within some pre-determined deadtime are assumed to be corrupted by pulse distortions. These are rejected, and the rest are accepted. This method can lower system jitter and maintain high data rates, especially in the cases where only a few percent of pulses are filtered out. However, it can severely limit count rate near the $3~\mathrm{dB}$ point where the majority of counts are rejected (see the supplementary material).

The calibration and correction approach we use requires no new hardware and preserves the original count rate of the detector. We calculate the time between a given \textit{current} SNSPD detection event and a preceding event. This inter-arrival time is used to determine a timing correction for the current event using a look-up table. A calibration routine described next is needed build this look-up table. Applying these corrections during real-time processing removes deterministic delays correlated with the time between time tags, leaving only stochastic jitter.

\begin{figure}[!htbp]
  \centering
  \includegraphics[width=1\linewidth]{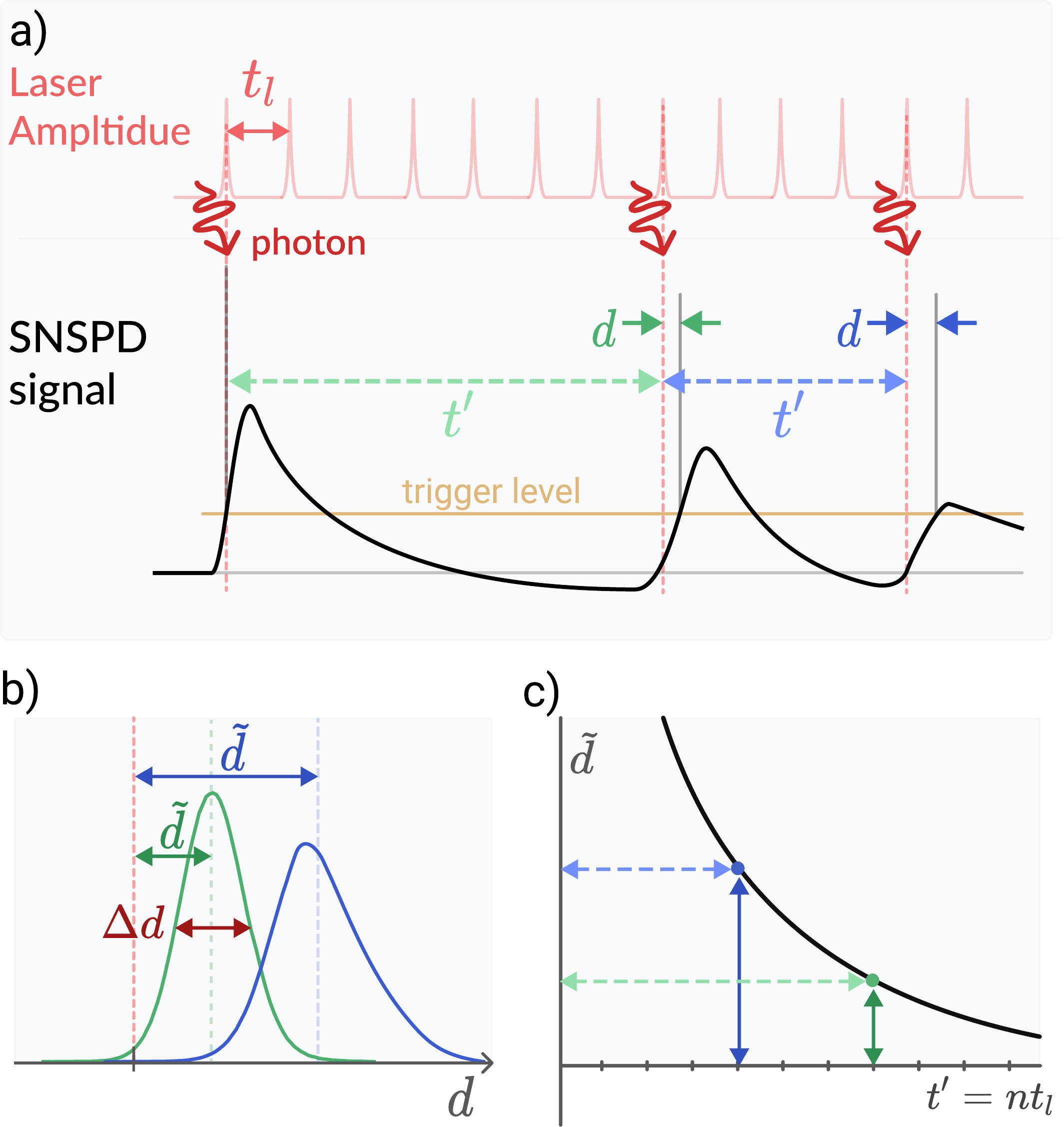}
  \caption{
    a) A qualitative diagram illustrating how inter-pulse timing measurements $t'$ and $d$ are extracted. A small fraction of laser pulses contain a photon due to the low mean photon number per pulse of the attenuated laser. Pairs of subsequent photon arrivals are separated by a time denoted by $t' = n t_l$. b) Possible distributions of delay $d$ measurements for two different $t'$. The median of these defines the extracted delay parameters $\tilde{d}$ which form the y-axis in the calibration curve illustrated in (c). The $\tilde{d}$ vs $t^\prime$ curve in (c) approaches zero for $t^\prime$ approaching infinity. Blue and green arrows with matching color and style denote the same measure in (a), (b), and (c). 
  }

  \label{fig:explain}
\end{figure}

We study the pulse distortions observed in a fiber coupled single-pixel tungsten silicide (WSi) SNSPD with 380~$\mathrm{\mu m^2}$ active area and 160~nm nanowire width. The detector is biased at 9.3~$\mathrm{\mu A}$, roughly 90\% of the switching current. The readout is handled by a cryogenic DC-coupled amplifier, mounted on the 40~K-stage of the cryostat, which has 43~dB of gain and a 3~dB bandwidth of 700 MHz, followed by a 1~GHz amplifier with 20~dB of gain (\textit{MiniCircuits ZFL-1000LN+}). The system reaches a $3~\mathrm{dB}$ maximum count rate (MCR) of $15.6~\mathrm{MHz}$. The time constant of supercurrent recovery $\tau_{bias} \approx 40~\mathrm{ns}$ is significantly longer than the decay time constant of the RF pulse $\tau_{RF} \approx 5~\mathrm{ns}$, owing to our use of a low-frequency cut-on of $\approx 80$~MHz (-3~dB point from the peak gain) in the DC-coupled amplifiers. For this detector, the calibration process primarily corrects for the lower bias current effects, rather than for any overlapping between RF pulse waveforms (see supplementary material). Other detector types and readout systems may operate in a different regime. 

The jitter increase with count rate is highly dependent on trigger level. High rate distortions affect the timing measurements less if the threshold voltage is set just above the noise floor. 
But triggering on the pulse higher, where it achieves maximum slope, minimizes jitter at low-to-medium count rates. This level varies from 20\% to 50\% of pulse height depending on the detector. We found the minimum low-rate jitter at a trigger level of 50~mV, about 40\% of the pulse height. All further calibration and analysis is performed by triggering at this level in order to optimize jitter across all count rates.

To perform our calibration, we illuminated the SNSPD with an attenuated 537.5~MHz mode locked laser with a mean photon number per pulse between $5\mathrm{e}{-4}$ and $0.016$. The 1.86~ns period of the pulse sequence is large enough that almost all SNSPD detections can be unambiguously matched with a preceding laser pulse -- the period of the pulse train used must be greater than the worst detector jitter for this to succeed. The uncorrected jitter for the WSi detector varies from 50~ps FWHM at low count rates, to about 350~ps at high rates.

We collect sorted time-tags and first consider adjacent pairs of SNSPD events as illustrated in Fig. \ref{fig:explain}a. The time between the two photons that produced these event pairs is defined as $t^\prime$, which is an integer number of laser periods ($t^\prime = n t_{l}$). Second, we derive the delay between the second event and its corresponding laser pulse, defined as $d$. For each laser period spacing $t^\prime$, we make a histogram of the second event delays and find the median ($\tilde{d}$) and the FWHM ($\Delta {d}$) of this distribution. For shorter separations $t^\prime$, the distribution is expected to have larger delays and FWHM (Fig.~\ref{fig:explain}b) due to the smaller pulse height of the second event. Finally, we use the median delay as a function of laser spacing (Fig.~\ref{fig:explain}c) to form a curve $\tilde{d}$-vs-$t^\prime$ for the added time-walk versus inter-arrival time.

\begin{figure}[!htbp]
  \centering
  \includegraphics[width=1\linewidth]{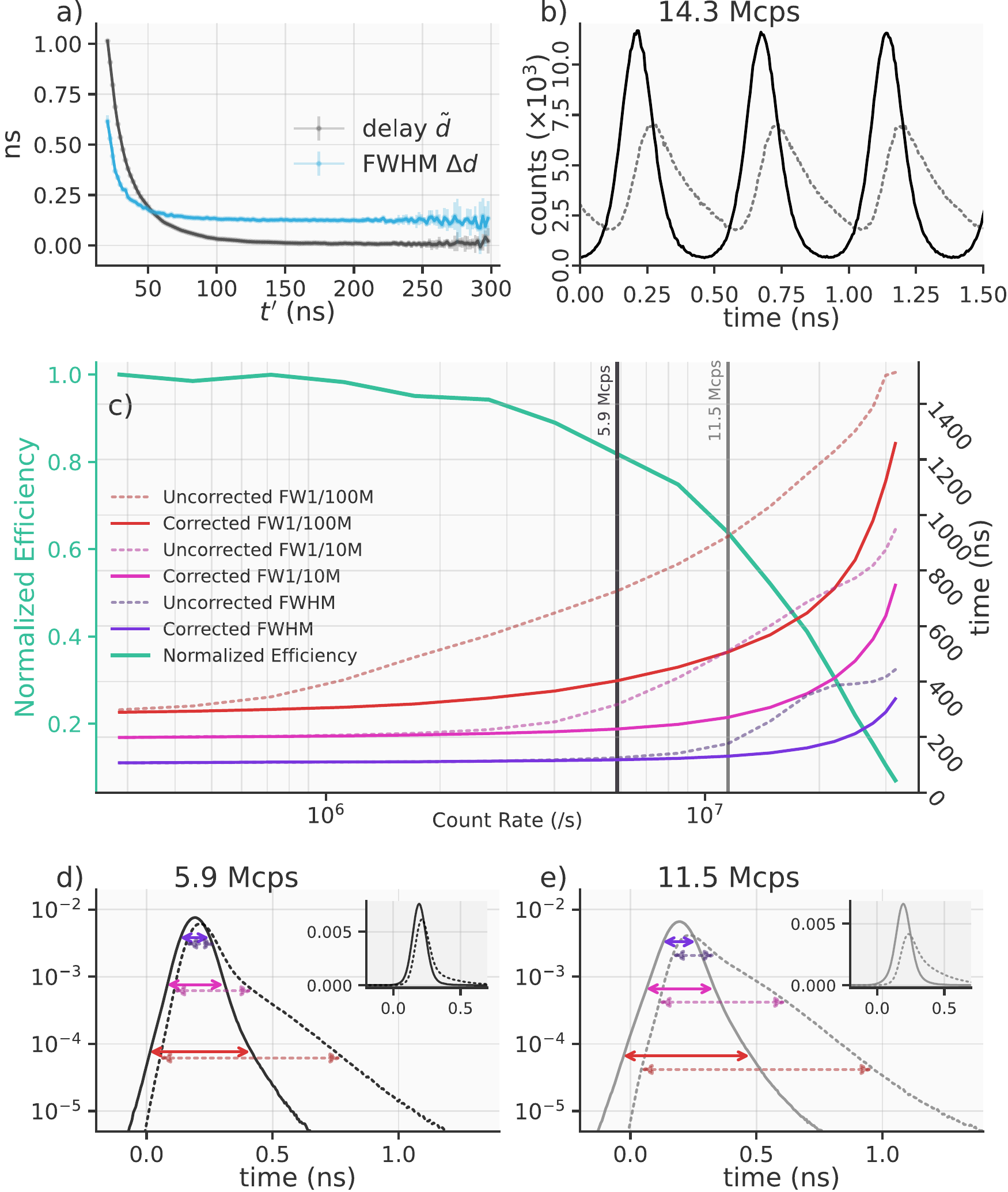}

  \caption{
    a) Delay and intrinsic jitter curves extracted from the 537.5 MHz pulsed light dataset. b) Histogram of corrected (black) and uncorrected (dashed grey) time tags from a 2.15 GHz pulse train, corrected using a calibration curve developed with the 537.5 MHz dataset. The improvement affirms that the light modulation used for an application need not match the repetition rate of the calibration laser. c) Effect of the correction on measurements of system jitter over a range of rates approaching the maximum count rate. d) Corrected (solid) and uncorrected (dashed) instrument response functions with color-matched arrows showing the location of the width-statistics plotted above in (c). The black vertical line in (c) is drawn at the count rate of this plot. Inset shows linear scaling. e) Corrected (solid) and uncorrected (dashed) histogram analogous to (d), but at a higher count rate indicated by the grey line in (c).
  }

  \label{fig:data}
\end{figure}

Figure~\ref{fig:data}a shows the $\tilde{d}$-vs-$t^\prime$ and $\Delta d$-vs-$t^\prime$ curves collected from our measurements of the WSi detector. The $\tilde{d}$-vs-$t^\prime$ curve is the main result of the calibration process and is used as a look-up table in the correction method. $\Delta d$ is a measure of the more intrinsic jitter that the correction method cannot cancel out. While it is larger for small $t^\prime$ due to triggers on lower amplitude pulses, it notably stays at a nearly minimized value down to around $t^\prime =$ 50~ns. $\tilde{d}$ grows more dramatically with decreasing $t^\prime$, especially in the range from 50~ to 100~ns. For count rates that don't exhibit many inter-pulse arrival times smaller than 50~ns, a method for removing the time-walk effect's contribution to jitter should bring entire system jitter back down to near the intrinsic limit implied by the $\Delta d$ curve.

The correction method we implement involves subtracting off the distortion-induced delays a time tag is expected to have based on the inter-pulse-time that precedes it. For each time tag $t_n$ in a set, the inter-pulse time is $t_n - t_{n-1} = \Delta t$, where $t_{n-1}$ is the previous tag on the same channel. Using $\Delta t$ as an index, a corresponding delay correction $\hat{d}$ is found by interpolating the $\tilde{d}$-vs-$t^\prime$ curve from the calibration. An assumption underlying the correction is that the interpolated value $\hat{d}$ is a good estimator of the true delay added to the current time tag due to high count rate pulse distortions.

With the interpolation operation expressed as a function $D$, the correction is written as $\overline{t}_n = t_n - D(\Delta t)$, where $\overline{t}_n$ is the corrected version of tag $t_n$. Whether $t_{n-1}$ is itself corrected or uncorrected has negligible influence on the $t_n$ correction, as we assume $d << \Delta t$. Therefore the correction process is separable and straightforward to implement as a real-time processing step in the FPGA or computer used for time tagging. 

The correction we perform using the $\tilde{d}$-vs-$t^\prime$ curve in Fig. \ref{fig:data}a is applicable to a wide range of count rates and arbitrary modulation patterns; there is no requirement that applications match the repetition rate of the calibration laser. Figure~\ref{fig:data}b shows that the correction method, derived from the 537.5 MHz calibration data, significantly reduces jitter when applied to detections from a 2.15~GHz pulse train. A similar jitter reduction can be demonstrated for repetition rates below 537.5~MHz. 

To study the effectiveness of our correction method at different count rates, we apply it to data collected at different mean photon numbers per pulse, with the same 537.5 MHz pulse train. As shown in Fig. \ref{fig:data}c, the correction improves the FWHM at rates approaching the 3~dB point, and improves FW10\%M and FW1\%M (full width at ten percent/one percent maximum) dramatically, even at count rates significantly below the 3 dB point, where detector efficiency is nearly maximized. This reduction is evident in \ref{fig:data}d, where the correction works to remove a time-walk induced tail in the instrument response function. The ratio of corrected FW1\%M over uncorrected FW1\%M reaches a minimum of 0.55 at a count rate of 11.5 MCounts/. Therefore if an application sets its repetition rate or bin size based on the FW1\%M metric, the repetition rate can be increased and the bin size decreased by up to 45\% without any increase in event misattribution errors. These improvements are notable for applications including biomedical imaging~\cite{Sutin16, Bruschini2019}, quantum communication~\cite{Hadfield2009} and laser ranging~\cite{McCarthy13} that have stringent timing requirements over a large dynamic range. 

For some SNSPD systems, the intrinsic reset time of the nanowire is considerably shorter than the reset dynamics of the amplifier chain. Then, the delay effect induced on each pulse may depend on the arrival time of multiple previous pulses, as amplifier reset dynamics combine additively. To optimally correct for this, higher-order correction techniques are needed based on higher-dimensional lookup tables. There is an avenue for exploring such methods for unique use-cases. However, the single-$\Delta t$ measurement approach detailed here is broadly applicable and straightforward to implement. 

As applications like LIDAR and quantum communication demand ever higher data rates, multiple techniques for increasing photon and data throughput of SNSPD systems are being explored. Arrays or multi-channel SNSPD systems will play a role in satisfying that demand. However, compared to multiple lower count rate SNSPDs operating in parallel, a single detector operating at high rate has certain advantages. First, it makes more efficient use of the extensive bandwidth of the RF readout channel. Second, the single detector with single readout line puts less thermal load on the cooling system than multiple detectors with multiple readout lines. Therefore, paths toward operating individual SNSPDs at the limits of their count rate performance should be explored before extending to multi-pixel systems. This work is a step towards unlocking all available performance and timing precision of SNSPDs operated at high count rates.

\section*{Acknowledgments} Part of the research was carried out at the Jet Propulsion Laboratory, California Institute of Technology, under a contract with the National Aeronautics and Space Administration (NASA) (80NM0018D0004). Support for this work was provided in part by the Defense Advanced Research Projects Agency (DARPA) Defense Sciences Office (DSO) Invisible Headlights program, NASA SCaN, Alliance for Quantum Technologies’ (AQT) Intelligent Quantum Networks and Technologies (INQNET) program and the Caltech/JPL PDRDF program.  A. M. is supported in part by the Brinson Foundation and the Fermilab Quantum Institute. M.S. is in part supported by the Department of Energy under grants SC0019219 and SC002376.  We are grateful to  Si Xie (Caltech/Fermilab) and Cristian Pena (Fermilab) for supporting this work in terms of sharing equipment and facilities. The authors are also grateful to Ioana Craiciu for her attention and help in editing the final manuscript.

\section*{Data Availability Statement} The data that support the findings of this study are openly available in SNSPD-time-walk-and-jitter-correction at \href{https://doi.org/10.6084/m9.figshare.20372646.v1}{https://doi.org/10.6084/m9.figshare.20372646.v1}.

\nocite{*}
\bibliography{references}

\end{document}


\maketitle
\section{Experimental Apparatus}
\begin{figure}
    \centering
    \includegraphics[width=0.7\linewidth]{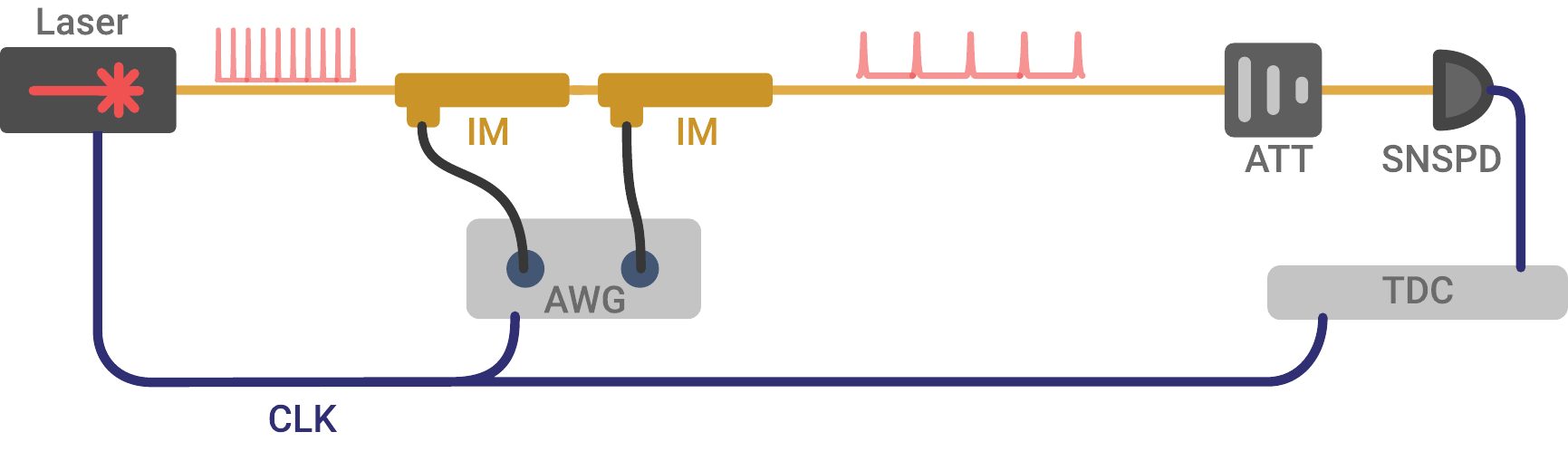}
    \caption{CLK: clock synchronization signal; AWG: Arbitrary Waveform Generator (\textit{Keysight}); IM: Intensity Modulator (\textit{IXBlue}); TDC: Time to Digital Converter (\textit{Swabian Instruments}). The extinction ratio of both modulators exceeds 30 dB. }
    \label{fig:expiremental_setup}
\end{figure}

The measurements used for calibration and correction were acquired with a high rate mode locked laser (\textit{Pritel}). The minimum repetition rate of this laser was too high to produce a calibration dataset for the Tungsten Silicide detector at highest count rate, given the constraints addressed in the main text. Therefore, the laser was set at a repetition rate of 10.75 GHz and modulated down to 537.5 MHz using two lithium niobate intensity modulators in series. Residual peaks from suppressed mode-locked laser pulses are not observed above the noise floor in the collected data, so their effect was not further considered. Clock jitter is minimized in post-processing with the help of a software-based Phase Locked Loop.

\section{Performance Regime of the Tungsten Silicide detector}

When our calibration and correction method is applied to the Tungsten Silicide SNSPD, the walk-cancellation method primarily corrects for pulse height variations. These variations are caused by varying levels of bias current in the device at the time of photo-detection. An oscilloscope trace shows an exponentially decaying increase in SNSPD pulse height following a previous detection. This exponential recovery shape reinforces evidence that this detector operates in this 'bias current recovery' regime, rather than the regime where amplifier reset dynamics dominate. 

\begin{figure}
    \centering
    \includegraphics[width=0.7\linewidth]{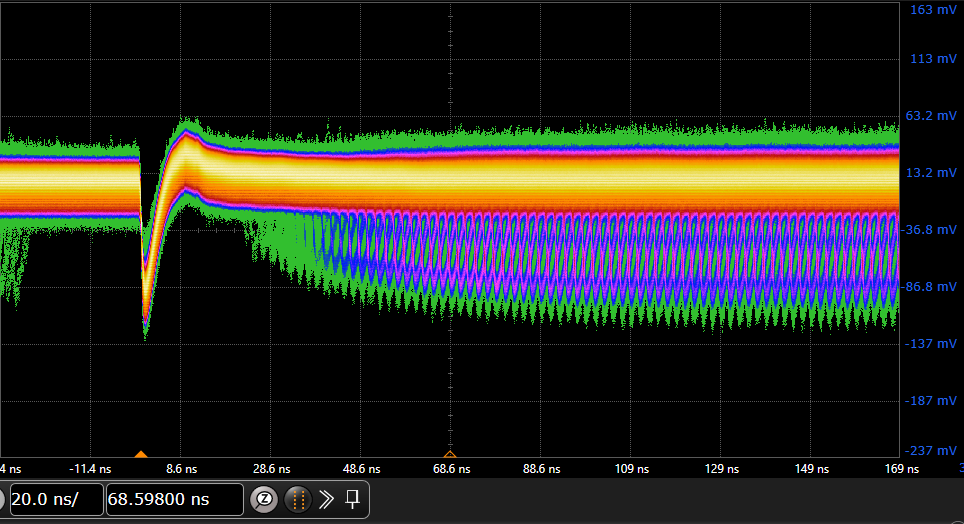}
    \caption{Scope trace of WSi detector illuminated by 537.5 MHz pulse train}
    \label{fig:scope}
\end{figure}

\section{Software dead time for high count rate jitter suppression}
\begin{figure}
    \centering
    \includegraphics[width=0.7\linewidth]{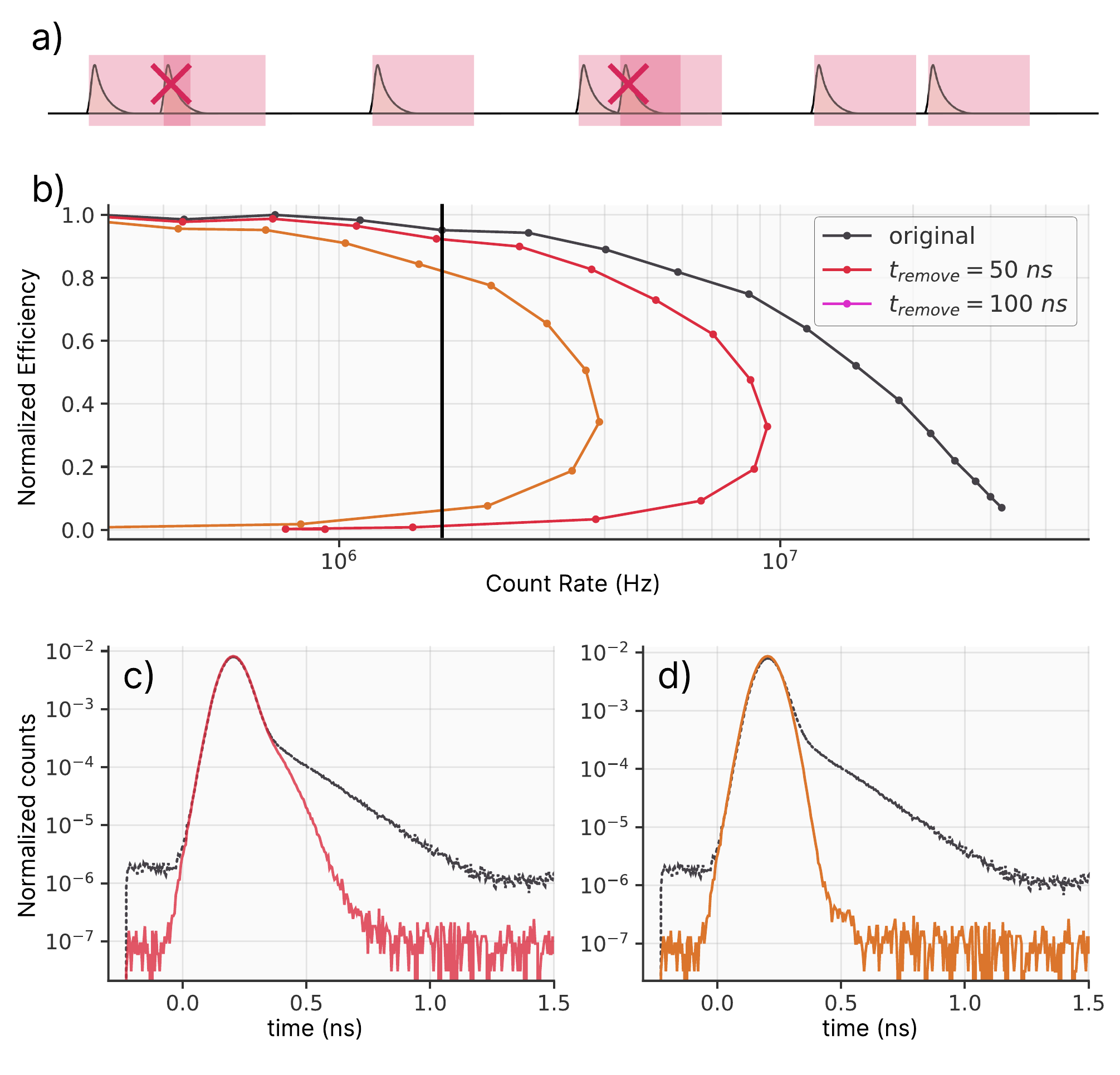}
    \label{fig:dead_time}
    \caption{a) Illustration of the RF signal out of an SNSPD with red highlighted regions signifying a software-based dead time that rejects some events. b) Count rate vs normalized efficiency, similar to Fig. 3c in green in the main text. c) \& d) Response functions for filtered and unfiltered timetags, similar to Fig. 3d and 3e from the main text. }
\end{figure}

In certain cases a software-based dead time is an effective way of reducing jitter at high count rates. SNSPD pulses that arrive soon after a previous pulse are ignored because their timing is assumed to be corrupted due to pulse distortions (Fig. \ref{fig:dead_time}a). With a long software-based dead time, data is filtered to keep only events for which the SNSPD was in a fully reset state prior to detection. This results in low jitter measurements even at high rate as shown in Fig. \ref{fig:dead_time}d where the dashed and solid lines are response functions of unfiltered and filtered data respectively. However, the use of software-based  dead times can severely limit usable count rate. This paradoxically contrasts with the main intended goal, which is to operate an SNSPD at the highest possible count rates. As shown in the Fig. 3b, adding a 100 ns software dead time to our WSi single pixel SNSPD limits its usable maximum count rate to about 4 MHz, while the raw count rate exceeds 10 MHz. Furthermore, the usable count rate drops to zero for higher incident photon rates, as the dead time starts to reject most events. This behavior can be unexpected and problematic for any applications that occasionally over-saturate the detector.